\documentclass[magazine]{IEEEtran}
\usepackage{cite}
\usepackage{amsmath,amssymb,amsfonts}
\usepackage{graphicx}
\usepackage{multirow}
\usepackage{subcaption}
\usepackage{caption}
\usepackage{float}
\usepackage{textcomp}
\usepackage[dvipsnames]{xcolor}
\usepackage{tikz}
\usepackage{pgfplots}
\usepackage{stfloats}
\usepackage{mathrsfs}
\usepackage{balance}
\usepackage{amsthm} 
\usepackage[ruled,vlined]{algorithm2e}

\usepackage{comment}
\usepackage{algpseudocode}
\usepackage{acronym}
\usepackage[hyperfirst=true]{glossaries}
\usepackage[hidelinks]{hyperref} 
\usepackage{makecell}
\makeglossaries
\newacronym{OFDM}{OFDM}{orthogonal frequency division multiplexing}
\newacronym{OTFS}{OTFS}{orthogonal time frequency space}
\newacronym{SNR}{SNR}{signal-to-noise ratio}
\newacronym{AoA}{AoA}{angle of arrival}
\newacronym{BER}{BER}{bit error rate}
\newacronym{DL}{DL}{deep learning}
\newacronym{FNN}{FNN}{feedforward neural network}
\newacronym{ICI}{ICI}{intercarrier interference}
\newacronym{TM}{TM}{threshold method}
\newacronym{IMFC}{IMFC}{iterative matched filtering and combining}
\newacronym{ETSI}{ETSI}{European Telecommunications Standards Institute}
\newacronym{3GPP}{3GPP}{Third Generation Partnership Project}
\newacronym{RE}{RE}{resource element}
\newacronym{ISAC}{ISAC}{integrated sensing and communications}
\newacronym{DD}{DD}{delay-Doppler}
\newacronym{OOB}{OOB}{out-of-band}
\newacronym{TF}{TF}{time-frequency}
\newacronym{SFFT}{SFFT}{symplectic finite Fourier transform}
\newacronym{CSF}{CSF}{channel spreading function}
\newacronym{MMSE}{MMSE}{minimum mean squared error}
\newacronym{CRLB}{CRLB}{Cramér-Rao lower bound}
\newacronym{PAPR}{PAPR}{peak-to-average power ratio}
\newacronym{6G}{6G}{sixth-generation}
\newacronym{HSR}{HSR}{high-speed railways}
\newacronym{V2V}{V2V}{vehicle-to-vehicle}
\newacronym{V2I}{V2I}{vehicle-to-infrastructure}
\newacronym{V2X}{V2X}{vehicle-to-everything}
\newacronym{UAV}{UAV}{unmanned aerial vehicle}
\newacronym{GS-OFDM}{GS-OFDM}{gear-switching OFDM}
\newacronym{BS}{BS}{base station}
\newacronym{4G}{4G}{fourth-generation}
\newacronym{5G}{5G}{fifth-generation}
\newacronym{LTE}{LTE}{long-term evolution}
\newacronym{CP}{CP}{cyclic prefix}
\newacronym{SC-FDMA}{SC-FDMA}{single-carrier frequency division multiple access}
\newacronym{MIMO}{MIMO}{multiple-input multiple-output}
\newacronym{NR}{NR}{new radio}
\newacronym{mmWave}{mmWave}{millimeter-wave}
\newacronym{LEO}{LEO}{low-Earth orbit}
\newacronym{RAN}{RAN}{radio access network}
\newacronym{DD-a-OFDM}{DD-a-OFDM}{delay-Doppler domain signal processing aided OFDM}
\newacronym{DDW-OFDM}{DDW-OFDM}{dual-domain waveform OFDM}
\newacronym{UE}{UE}{user equipment}
\newacronym{CFO}{CFO}{carrier frequency offset}
\newacronym{LoS}{LoS}{line-of-sight}
\newacronym{FFT}{FFT}{fast Fourier transform}
\newacronym{QoS}{QoS}{quality of service}
\newacronym{C-V2X}{C-V2X}{cellular V2X}
\newacronym{DCI}{DCI}{downlink control information}
\newacronym{RRC}{RRC}{radio resource control}
\newacronym{CSI}{CSI}{channel state information}
\newacronym{DMRS}{DMRS}{demodulation reference signal}
\newacronym{LS}{LS}{least squares}
\newacronym{CTF}{CTF}{channel transfer function}
\newacronym{CE}{CE}{channel estimation}
\newacronym{SP}{SP}{superimposed pilot}
\newacronym{PDR}{PDR}{pilot-to-data ratio}
\newacronym{PDP}{PDP}{power delay profile}
\newacronym{CCDF}{CCDF}{complementary cumulative distribution function}
\newacronym{SCS}{SCS}{subcarrier spacing}
\usepackage{orcidlink}
\usepackage{tabularx}   
\usepackage{booktabs}   
\usepackage{array}
\usepackage{colortbl}
\usepackage[dvipsnames]{xcolor}

\definecolor{gear1green}{RGB}{76,175,80}
\definecolor{gear2yellow}{RGB}{255,193,7}
\definecolor{gear3red}{RGB}{244,67,54}
\definecolor{chalmersblue}{RGB}{0,73,131}
\definecolor{lightgray}{RGB}{245,245,245}
\definecolor{lightblue}{RGB}{227,242,253}
\definecolor{lightyellow}{RGB}{255,249,196}
\definecolor{lightred}{RGB}{255,235,238}

\title{Beyond Legacy OFDM: A Mobility-Adaptive Multi-Gear Framework for 6G}

\author{Mauro~Marchese$^{\orcidlink{0009-0008-0265-5840}}$,~\IEEEmembership{Graduate Student Member,~IEEE}, Dario~Tagliaferri$^{\orcidlink{0000-0002-5718-4571}}$,~\IEEEmembership{Member,~IEEE},\\[ -1.2ex] Henk~Wymeersch$^{\orcidlink{0000-0002-1298-6159}}$,~\IEEEmembership{Fellow,~IEEE}, 
Musa~Furkan~Keskin$^{\orcidlink{0000-0002-7718-8377}}$,~\IEEEmembership{Member,~IEEE},
\\[ -1.2ex]Emanuele~Viterbo$^{\orcidlink{0000-0002-5861-2873}}$,~\IEEEmembership{Fellow,~IEEE}, and Pietro~Savazzi$^{\orcidlink{0000-0003-0692-8566}}$,~\IEEEmembership{Senior Member,~IEEE}
\thanks{Mauro~Marchese and Pietro~Savazzi are with the Department of Electrical, Computer and Biomedical Engineering, University of Pavia, 27100 Pavia, Italy.}
\thanks{Musa~Furkan~Keskin and Henk~Wymeersch are with the Department of Electrical Engineering, Chalmers University of Technology, 41296 Gothenburg, Sweden.}
\thanks{Dario~Tagliaferri is with the Department of Electronics, Information and Bioengineering, Politecnico di Milano, 20133 Milan, Italy.}
\thanks{Emanuele~Viterbo is with the Department of Electrical and Computer Systems Engineering, Monash University, Clayton, VIC 3800, Australia.}
\thanks{This work is supported, in part, by grants from the Swedish Research Council (no. 2022-03007 and 2024-04390).}
}

\begin{document}
\maketitle

%TC:ignore
\begin{abstract}
While \gls{3GPP} has confirmed \gls{OFDM} as the baseline waveform for \gls{6G}, its performance is severely compromised in the high-mobility scenarios envisioned for \gls{6G}. 
Building upon the GEARBOX-PHY vision, we present \textit{\gls{GS-OFDM}}: a unified framework in which the \gls{BS} adaptively selects among three gears, ranging from legacy \gls{OFDM} to delay-Doppler domain processing based on the channel mobility conditions experienced by the \glspl{UE}.
We illustrate the benefit of adaptive gear switching for communication throughput and, finally, we conclude  with an outlook on research challenges and opportunities.
\end{abstract}
%TC:endignore


\glsresetall

\section{Introduction}
\label{sec:intro}
%% ====================================================================

The transition to \gls{4G} \gls{LTE} marked the widespread adoption of \gls{OFDM} for the downlink, paired with \gls{SC-FDMA} for the uplink. \gls{OFDM} became the dominant waveform thanks to a compelling combination of properties: robustness to frequency-selective fading via the \gls{CP}, low-complexity single-tap equalization in the frequency domain, flexible time-frequency resource allocation, and a natural fit with \gls{MIMO} spatial multiplexing. \Gls{5G} \gls{NR} retained \gls{OFDM} with extensions including flexible numerology %---variable \glspl{SCS} from 15~kHz up to 240~kHz---
to accommodate diverse deployment scenarios from sub-6\,GHz to \gls{mmWave} bands \cite{parkvall2018nr}.

The journey toward \gls{6G} systems raises new challenges \cite{uusitalo2025toward}. On the  \textit{communication} side, \gls{6G} is expected to support high-mobility use cases far beyond the scope of \gls{5G}: \gls{HSR} at velocities up to 500~km/h, low-altitude \gls{UAV} networks, \gls{V2X} communications on highways, and connectivity to \gls{LEO} satellites. At these velocities, the Doppler effect induces significant \gls{ICI}, breaking the subcarrier orthogonality that underpins \gls{OFDM}'s low-complexity equalization \cite{Viterbo2022}. The channel becomes doubly selective (varying in both time and frequency) and the channel coherence time and bandwidth reduce.
Increasing the \gls{SCS}   shortens the \gls{OFDM} symbol duration and improves robustness to Doppler spread, but it simultaneously increases the \gls{CP} overhead (degrading spectral efficiency) or limits the maximum tolerable delay spread. This intrinsic \gls{SCS} trade-off makes numerology tuning alone insufficient to address the wide range of mobility conditions envisioned for \gls{6G}.

On the \textit{sensing} side, \gls{6G} is the first generation of wireless networks that is envisioned to natively support \gls{ISAC}, namely the intertwined fusion of data transmission and environment perception, from basic target localization to full 2D/3D environment mapping, for a plethora of applications, such as autonomous mobility, public safety, and others~\cite{GonzalezPrelcic2024,etsi_gr_isc_001_v1_1_1_2025}. This dual functionality places new demands on the waveform: beyond throughput and \gls{BER}, this shall enable high-accuracy and high-resolution sensing, requiring good \gls{DD} ambiguity properties that are not straightforwardly achievable with legacy waveforms (\gls{OFDM}). While standardization bodies had recently initiated \gls{ISAC}-specific efforts, such as \gls{ETSI} \cite{etsi_gr_isc_003_v1_1_1_2026} and \gls{3GPP} (with Release~19 \cite{3gpp_release19_workplan}), signaling the importance of this capability for the \gls{6G} ecosystem, no clear answer has been provided about what waveform is best suited for \gls{ISAC}. 

%On the \textit{sensing} side, \gls{6G} is envisioned to natively provide \gls{ISAC}, enabling the radio access network to simultaneously transmit data and perceive the environment~\cite{GonzalezPrelcic2024}. This dual functionality places new demands on the waveform: beyond throughput and \gls{BER}, the waveform must support high-resolution delay-Doppler estimation, favorable ambiguity function properties, and the ability to extract physical scatterer parameters. The 3GPP standardization body has initiated \gls{ISAC} channel modeling studies in Release~19, signaling the importance of this capability for the \gls{6G} ecosystem.

These challenges have motivated extensive research into alternative waveforms that operate in the \gls{DD} domain, as it is native to radar sensing applications \cite{Mohammed2022} and exhibits a sparse and quasi-static structure \cite{Viterbo2022}. In \cite{Hadani2017}, \gls{OTFS} modulation has been introduced, pioneering the concept of multiplexing data symbols on a \gls{DD} grid, exploiting the properties of the \gls{DD} channel. While these waveforms offer superior reliability in high-Doppler environments, they face practical barriers: \gls{OTFS}'s \gls{DD}-domain multiplexing is less flexible compared to time-frequency resource allocation, and receiver's processing latency in the full \gls{DD}-domain scales with the frame size rather than at a symbol-wise level. Transitioning away from \gls{OFDM} would entail significant redesigns of existing radio hardware and software frameworks.

In August 2025, the \gls{3GPP} \gls{RAN} working groups agreed that \gls{CP}-\gls{OFDM} and \gls{SC-FDMA} will serve as the baseline waveforms for \gls{6G} in Release~20 \cite{3gpp_ran1_122_report}. Crucially, this decision explicitly leaves the door open for enhancements and modifications to \gls{CP}-\gls{OFDM}, and does not preclude other \gls{OFDM}-based waveforms.

\begin{figure}[t]
\centering
    \includegraphics[width=\linewidth, trim=80 80 80 20, clip]{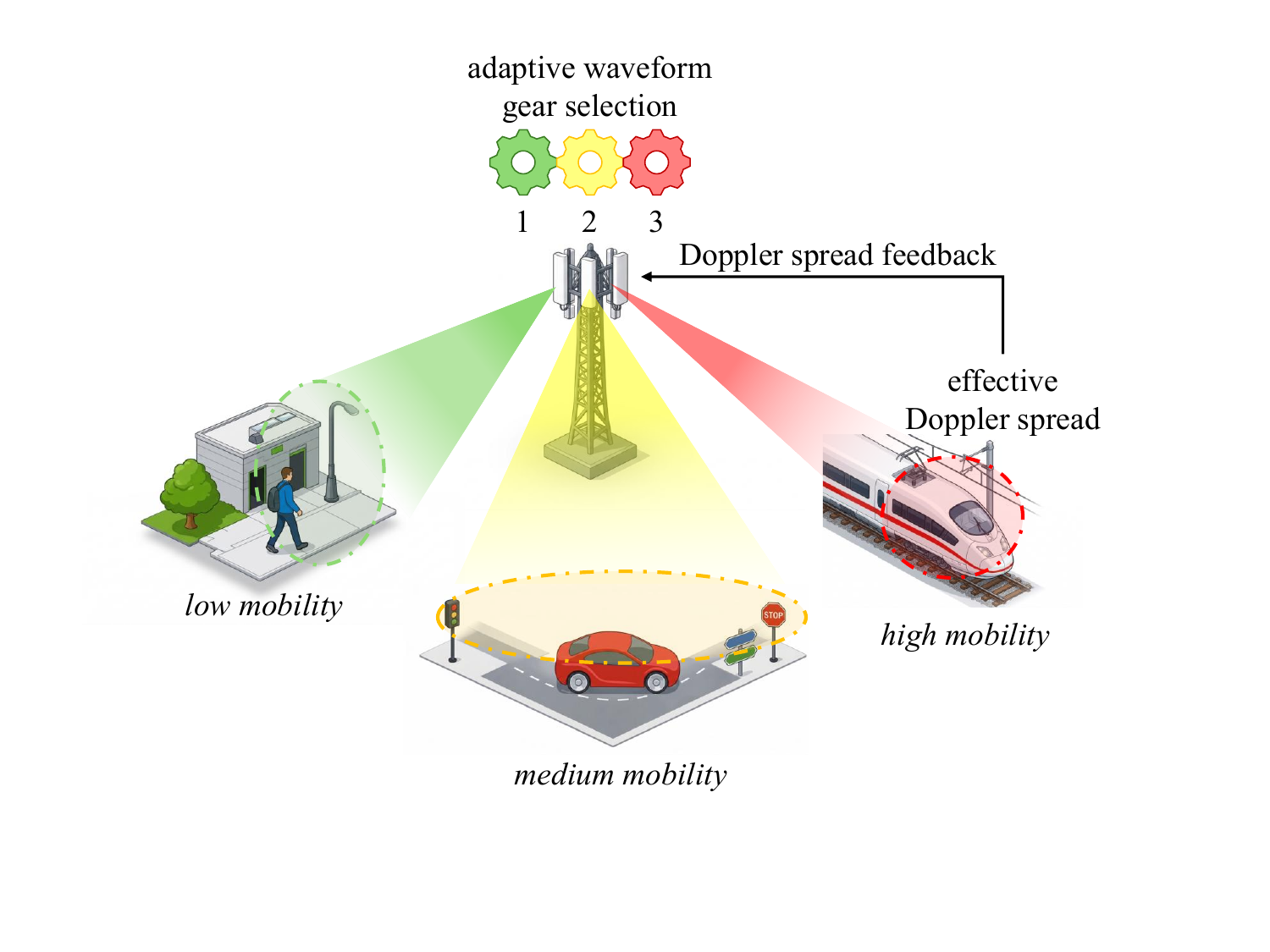}
    \caption{The \gls{GS-OFDM} framework: the \gls{BS} adaptively selects among three operating gears based on the mobility conditions and Doppler spread feedback from the \glspl{UE}. Gear 1 is used for low mobility (quasi-static), Gear 2 for medium mobility (time-varying), and Gear 3 for high/extreme mobility (fast-varying) scenarios.}\label{fig:gear_concept}
\end{figure}

While \gls{OFDM} is the 6G waveform, the way it is used (e.g., pilot design, precoding, receiver processing) is open for innovation. %It is precisely this opportunity that we target in this work, leveraging the concept of GEARBOX-PHY introduced and investigated for next-generation wireless systems in \cite{Fettewis2021,Gast2024}. 
Specifically, at the physical layer, the modulating waveform could be properly selected among different ``waveform gears", depending on the specific channel conditions, in line with the  GEARBOX-PHY vision from  \cite{Fettewis2021}.
%In this paper, 
To this end, we present \textit{\gls{GS-OFDM}}: a unified framework in which the \gls{BS} adaptively selects among three operating gears depending on the channel mobility conditions, as illustrated in Fig.~\ref{fig:gear_concept}. %The metaphor is that of a \emph{waveform gear}: just as a vehicle uses low gear for steep inclines and high gear for highways, the \gls{BS} shifts to a higher processing gear when the Doppler environment becomes more challenging. 
The three gears encompass: legacy \gls{OFDM} (Gear~1); \gls{DD-a-OFDM} with delay-Doppler channel estimation from standard time-frequency pilots (Gear~2) \cite{Yiyan2025}, which primarily modifies the receiver processing; and \gls{DDW-OFDM}, which utilizes \gls{DD} superimposed pilot processing with iterative equalization (Gear~3) \cite{marchese2025robust6gofdmhighmobility,Tagliaferri2024}, thereby requiring changes to both transmitter and receiver architectures. The gear selection is driven by the estimated Doppler spread by the \gls{UE}. This information is fed back at the \gls{BS} via uplink transmission, and the \gls{BS} selects the proper waveform gear to serve the user.
By keeping the \gls{SCS} fixed and adapting only the signal processing gear, \gls{GS-OFDM} preserves delay spread tolerance and spectral efficiency while extending reliable operation to progressively higher Doppler spreads, a flexibility that numerology tuning alone cannot provide.

%TC:ignore
%TC:endignore

%% ====================================================================
\section{Basic Principles and Motivation}
\label{sec:channel}
%% ====================================================================

The wireless channel model is the physical foundation that determines when each processing gear is needed. In this section, we present basic knowledge on wireless channel modeling and we discuss the impact of mobility on the channel response. 

\subsection{Channel Model and the Impact of Mobility}
The wireless signal, when traveling from the transmitter to the receiver, bounces off multiple reflectors. These multipath components are each characterized by three parameters: a complex channel gain, a propagation delay, and a Doppler shift. This multipath channel point of view provides the very sparse \gls{DD}-domain channel model, in contrast with the \gls{TF}-domain channel model with a large number of parameters \cite{Hadani2017}, commonly adopted by legacy \gls{OFDM}. The key observation from the \gls{DD}-domain perspective is that, whereas the \gls{TF} domain channel parameters (per-subcarrier channel gains) vary rapidly in both dimensions under high mobility, the \gls{DD}-domain representation remains sparse and quasi-static---the delays and Doppler shifts remain approximately constant over time scales much longer than the \gls{OFDM} frame duration \cite{Viterbo2022}. This stability is the foundation of \gls{DD}-domain processing.
A careful distinction is required between two components of the Doppler effect: the \textit{common Doppler shift} and the \textit{Doppler spread}. The ego-motion of the \gls{UE} along the dominant propagation direction (e.g., the \gls{LoS} path) induces a common Doppler shift across all subcarriers. This component can be estimated and compensated at the receiver prior to the \gls{FFT}. After compensation, the common Doppler does not cause \gls{ICI}. On the other hand, the variation of Doppler shifts across paths, arising from scatterers at different angles and/or with different velocities, is the quantity that determines the effective channel time variation. The Doppler spread causes the channel to change within and across \gls{OFDM} symbols, leading to \gls{ICI} that cannot be removed by a single frequency correction. The channel coherence time is conversely proportional to the Doppler spread.

%\subsection{Motivation for Adaptive Gear-Switching}

%{\color{ForestGreen}We assume that the \gls{OFDM}-based system transmits frames of duration $T_f$ with $N$ OFDM symbols of duration $T+T_{\text{CP}}$ each. This signal will be sent through a wide range of time-varying channels with different parameter stability/stationarity time ranges, relative to the frame and the OFDM symbol durations.}
%The channel coherence time is inversely proportional to the Doppler spread. 
The number of \gls{OFDM} symbols during which the channel is approximately constant is given as the ratio of the channel coherence time and the \gls{OFDM} symbol duration, including the \gls{CP}.
\begin{comment}
\begin{equation}
N_{\text{c}} = \left\lfloor \frac{T_c}{T_{\text{sym}}} \right\rfloor = \left\lfloor \frac{1}{\sigma_{\nu} \cdot T_{\text{sym}}} \right\rfloor,
\label{eq:Ncoh}
\end{equation}
where $T_{\text{sym}} = 1/\Delta f + T_{\text{CP}}$ is the \gls{OFDM} symbol duration including the \gls{CP}. 
\end{comment}
%As this parameter quantifies the number of time-slots during which the channel remains approximately constant, it could be a good metric for the adaptive gear-switching protocol:
Based on that, we can identify three different channel conditions:
\begin{itemize}
\item Low mobility: the channel is quasi-static within the frame or varies slowly. In this case, \gls{ICI} is negligible as the Doppler spread is almost less than ten percent of the \gls{SCS} \cite{Sturm2011}. Conventional low-complexity \gls{OFDM} processing is enough to provide good \gls{QoS}.
\item Medium mobility: the channel varies noticeably across the frame and \gls{ICI} is no longer negligible. In this case, legacy \gls{OFDM} experiences performance degradation.
\item High mobility: the channel varies rapidly within a few symbols and \gls{ICI} is detrimental for legacy \gls{OFDM} implementation.
\end{itemize}

\subsection{Quantitative Assessment}
\label{sec:channel_quant}
To quantify the necessity for adaptive processing, we evaluate the effective number of coherent \gls{OFDM} symbols across a wide range of system configurations and velocities, representing a unified urban vehicular scenario. The results, reported in Table~\ref{tab:coherence}, highlight the impact of both the carrier frequency and the \gls{SCS} on channel coherence. These numbers are obtained considering the worst-case Doppler spread experienced by an isotropic single-antenna receiver under different system configurations and environmental conditions (see below for more insights) and serve as a representative rather than exhaustive example, demonstrating the extreme variability of channel conditions to justify the requirement for our proposed adaptive approach.

Several key observations emerge from the data. The \gls{C-V2X} configuration (5.9\,GHz with \gls{SCS} of $15$\,kHz) represents the most coherence-limited case; due to the relatively long symbol duration, the channel remains quasi-static only at low urban speeds. As mobility increases to highway levels ($\approx 120$\,km/h), the system rapidly enters the medium-mobility regime, requiring \gls{DD}-domain tracking even at relatively low carrier frequencies.
In the FR1 (3.5\,GHz) and FR3 (15\,GHz) bands, the adoption of larger \gls{SCS}s allows the system to maintain Gear~1 operation for a broader range of velocities. However, these configurations still transition through all three gears as mobility scales, reaching Gear~3 territory for high-speed rail or \gls{UAV} scenarios. Conversely, at FR2 and sub-THz frequencies, the wide \gls{SCS} (up to 480\,kHz) provides significant robustness. The resulting shortened symbol duration keeps these systems firmly in Gear~1 up to highway speeds, with Gear~3 required only under extreme conditions.

The results presented in Table~\ref{tab:coherence} provide a quantitative justification for a multi-gear waveform: \emph{channel conditions vary significantly depending on the experienced mobility and system configuration so that legacy \gls{OFDM} is not optimal across the entire operating space for 6G}. While legacy \gls{OFDM} is efficient for the green-shaded regions (low-mobility scenarios), it fails to maintain reliability in the red-shaded cells. Conversely, \gls{DD} processing is necessary for rapidly-varying channels but would introduce unnecessary computational overhead in quasi-static scenarios. 

%TC:ignore
\begin{table*}[!t]
\centering
\caption{Effective number of coherent OFDM symbols across velocities and system configurations. Colors indicate the channel conditions: \colorbox{green!10}{Low mobility}, \colorbox{yellow!10}{Medium mobility}, \colorbox{red!10}{High mobility}.}
\label{tab:coherence}
\small
\begin{tabular}{@{}l *{6}{c} @{}}
\toprule
& \textbf{C-V2X, 5.9\,GHz} & \textbf{FR1, 3.5\,GHz} & \textbf{FR3, 15\,GHz} & \textbf{FR3, 15\,GHz} & \textbf{FR2, 28\,GHz} & \textbf{Sub-THz, 140\,GHz} \\
\textbf{Velocity} & \gls{SCS}$=\!15$\,kHz & \gls{SCS}$=\!30$\,kHz & \gls{SCS}$=\!60$\,kHz & \gls{SCS}$=\!120$\,kHz &  \gls{SCS}$=\!120$\,kHz &  \gls{SCS}$=\!480$\,kHz \\
\midrule
3\,km/h (pedestrian)
  & \cellcolor{green!10}$>100$
  & \cellcolor{green!10}$>100$
  & \cellcolor{green!10}$>100$
  & \cellcolor{green!10}$>100$
  & \cellcolor{green!10}$>100$
  & \cellcolor{green!10}$>100$ \\
30\,km/h (urban)
  & \cellcolor{green!10}$\approx 45$
  & \cellcolor{green!10}$>100$
  & \cellcolor{green!10}$\approx 71$
  & \cellcolor{green!10}$>100$
  & \cellcolor{green!10}$\approx 77$
  & \cellcolor{green!10}$\approx 61$ \\
120\,km/h (highway)
  & \cellcolor{green!10}$\approx 11$
  & \cellcolor{green!10}$\approx 38$
  & \cellcolor{green!10}$\approx 17$
  & \cellcolor{green!10}$\approx 35$
  & \cellcolor{green!10}$\approx 19$
  & \cellcolor{green!10}$\approx 15$ \\
350\,km/h (HSR)
  & \cellcolor{yellow!10}$\approx 3$
  & \cellcolor{green!10}$\approx 13$
  & \cellcolor{yellow!10}$\approx 6$
  & \cellcolor{green!10}$\approx 12$
  & \cellcolor{yellow!10}$\approx 6$
  & \cellcolor{yellow!10}$\approx 5$ \\
500\,km/h (HSR,\ UAV)
  & \cellcolor{red!10}$\approx 2$
  & \cellcolor{yellow!10}$\approx 9$
  & \cellcolor{yellow!10}$\approx 4$
  & \cellcolor{yellow!10}$\approx 8$
  & \cellcolor{yellow!10}$\approx 4$
  & \cellcolor{yellow!10}$\approx 3$ \\
1000\,km/h (extreme)
  & \cellcolor{red!10}$\approx 1$
  & \cellcolor{yellow!10}$\approx 4$
  & \cellcolor{red!10}$\approx 2$
  & \cellcolor{yellow!10}$\approx 4$
  & \cellcolor{red!10}$\approx 2$
  & \cellcolor{red!10}$\approx 1$ \\
\bottomrule
\end{tabular}
\vspace{2pt}
\end{table*}
%TC:endignore

%% ====================================================================
\section{The Waveform Gear: GS-OFDM}
\label{sec:gear}
%% ====================================================================

Leveraging the quantitative insights discussed above, which motivate the need for an adaptive gear-switching protocol based on the number of coherent \gls{OFDM} symbols, this section presents the proposed waveform gear framework and describes the transition logic between the three processing modes.

\subsection{The Waveform Gear Concept}

No single \gls{OFDM} processing strategy is optimal across all mobility conditions. Legacy \gls{OFDM} with single-tap equalization is the simplest and most efficient at low mobility, but it fails when the Doppler spread causes significant \gls{ICI}. \gls{DD}-domain processing provides robustness at high mobility, but at the cost of increased complexity, higher latency (frame-based rather than symbol-based processing), and less flexibility. Using the most robust processing in any mobility condition wastes resources and introduces unnecessary overhead for the \glspl{UE} that are stationary or slow-moving.

The waveform gear resolves this tension through adaptive gear switching. The \gls{BS} acts as the intelligent controller as follows.
%\begin{enumerate}
%\item 
Each \gls{UE} estimates the \gls{CSI} of the downlink channel and feeds back to the \gls{BS} the estimate of the Doppler spread, or, more in general, the estimated Doppler spectrum, via uplink transmission. 
%\item
The \gls{BS} collects the Doppler spread experienced by each \gls{UE} and identifies clusters of \glspl{UE} that are subject to similar mobility conditions. 
%\item 
The \gls{BS} then selects the processing gear for each cluster of users according to proper thresholds on the estimated channel coherence time (derived from the Doppler spread).
%\item 
Finally, the gear selection is signaled to the \gls{UE}. %via \gls{DCI} for fast adaptation, or via \gls{RRC} signaling for slower, macro-mobility transitions. 
Optionally, the \gls{BS} can leverage the estimated Doppler spectrum by the \gls{UE} to \textit{pre-compensate} the corresponding spread and mitigate the \gls{ICI}. 
%\end{enumerate}
%
Hysteresis margins should be applied to the switching thresholds to avoid oscillation between gears due to noisy Doppler spread estimates. %The gear transitions are summarized as follows: Gear~1 $\to$ Gear~2 when $N_{\text{c}}$ drops below approximately $10$ symbols; Gear~2 $\to$ Gear~3 when $N_c$ drops below approximately $2$ symbols; reverse transitions occur with appropriate hysteresis. Importantly, gear transitions are not instantaneous: switching from Gear~1 to Gear~3 changes both the pilot scheme (requiring TX-side adaptation) and the receiver pipeline (from single-tap \gls{FFT} to iterative \gls{DD} equalization), causing a transient increase in processing latency and power consumption. 

The waveform gears within the proposed \gls{GS-OFDM} framework are detailed in the following sections.

\subsection{Gear~1: Legacy OFDM}
\label{sec:gear1}

%\textit{When:} 
%$N_{c} \gg 10$. 
When the channel is quasi-static or varies slowly within the \gls{OFDM} frame (corresponding to most low-mobility scenarios),  standard \gls{OFDM} with embedded time-frequency pilots following the \gls{5G} \gls{NR} \gls{DMRS} pattern is preferred. %No modifications to the transmitted waveform.
%
%\textit{Receiver (\gls{UE}):} 
The \gls{UE} applies
\gls{LS} or \gls{MMSE} channel estimation at pilot positions, followed by two-dimensional interpolation across the \gls{TF} grid, single-tap frequency-domain equalization, and standard demodulation/decoding.

%\textit{Properties:} 
Gear~1 offers the lowest complexity and latency (symbol-level processing) and full backward compatibility with \gls{5G} \gls{NR}. It is the default operating mode for the vast majority of \glspl{UE}.

\subsection{Gear~2: DD-a-OFDM}
\label{sec:gear2}

%\textit{When:} $2 \lesssim N_{c} \lesssim 10$. 
When the channel varies noticeably across the \gls{OFDM} frame (medium mobility scenarios), channel variations can be parameterized by a small number of \gls{DD}-domain parameters (delay, Doppler, complex gain per path). The system operates using \gls{DD-a-OFDM} \cite{Yiyan2025}, which consists of standard \gls{OFDM} with embedded \gls{TF} pilots at the \gls{BS} side. The pilot pattern can be slightly denser in time than in Gear~1 to support higher Doppler in the \gls{DD}-domain channel estimation processing. At the \gls{UE} side, the receiver exploits \gls{DD}-domain channel estimation as follows~\cite{Yiyan2025}: estimate the channel response at \gls{TF} pilot locations and obtain the downsampled \gls{CTF}, then convert the downsampled \gls{CTF} to the \gls{DD} domain and obtain the periodic \gls{CSF}. After that, channel parameters of each multipath component are estimated and the full time-varying \gls{TF}-domain channel matrices are reconstructed. The \gls{CE} is followed by \gls{MMSE} equalization in the \gls{TF} domain or via \gls{IMFC} \cite{marchese2025robust6gofdmhighmobility}.

Gear~2 adds moderate complexity at the receiver side (\gls{DD} parameter estimation) and introduces frame-based latency for channel estimation. However, it uses the standard \gls{TF} pilot structure, maintaining substantial backward compatibility. The limitation of this waveform gear is that the maximum Doppler spread that can be estimated and fed back to the BS is set by the minimum pilot spacing across time. Therefore, Gear~2 can not support reliable communications at extreme speeds without degrading the throughput, due to increased pilot density.

\subsection{Gear~3: DDW-OFDM}
\label{sec:gear3}

%\textit{When:} $N_c \lesssim 2$. 
When the channel varies rapidly (high mobility scenarios), \gls{TF}-domain pilots can sample the time variation only if the spacing is exactly one \gls{OFDM} symbol, drastically degrading the throughput. To overcome this issue, the \gls{GS-OFDM} framework operates using the \gls{DDW-OFDM} waveform proposed in \cite{marchese2025robust6gofdmhighmobility,Tagliaferri2024}. The \gls{BS} multiplexes \gls{OFDM} data to different users in the \gls{TF} domain and superimposes a \gls{DD}-domain pilot to the transmitted signal. The superimposed pilot is designed as either a single impulse or a more structured pattern in the \gls{DD} domain, transformed to the \gls{TF} domain and linearly superimposed on the data symbols with a controlled power ratio. The superimposed pilot employs all the available \gls{TF} resources (all the bandwidth and downlink burst duration). This allows channel estimation at the \gls{UE} side over the full \gls{DD} plane without the explicit need for \gls{TF} pilots.
%sacrificing data resource elements.
The \gls{UE} performs \gls{DD}-domain channel parameter estimation based on the superimposed pilot. After that, data symbols can be retrieved using, for instance, the \gls{IMFC} equalizer proposed in~\cite{marchese2025robust6gofdmhighmobility}, which decouples the equalization into per-path operations with computational complexity almost linear in the number of resource elements.

Gear~3 achieves the highest robustness but at the cost of the highest complexity and energy expenditure for the \gls{UE}, which requires the RF chain of the \gls{UE} to operate over the entire (or at least a fairly large) bandwidth, wider than the one that carries intended data symbols, to enable high-resolution \gls{DD}-domain processing paired with possibly iterative equalization. Another aspect to be considered is the effective power budget for channel estimation. The superimposed pilot, when transformed from the \gls{DD} domain to the \gls{TF} domain, spreads its energy across all subcarriers and \gls{OFDM} symbols. This raises the interference floor for the data payload. The \gls{PDR} \cite{marchese2025robust6gofdmhighmobility} is a critical design parameter: increasing the \gls{PDR} improves \gls{CE} accuracy (and thus equalization quality) but directly reduces the \gls{SNR} available for data detection. Gear~3 also requires changes to the pilot scheme, reducing backward compatibility with \gls{5G} \gls{NR}. It should be activated only in extreme mobility conditions.

\subsection{Sensing Considerations}

The \gls{GS-OFDM} framework opens interesting opportunities for sensing, both at the \gls{BS} and \gls{UE} side, which we shall review in the following.

\subsubsection{Gear 1}
Legacy \gls{OFDM} is the baseline for low-mobility scenarios. On the one hand, a \gls{BS} with full-duplex capabilities can sense the environment in monostatic configuration to detect the location of the main scattering centers in the environment and their radial velocity. The \gls{BS} processes the received signal over the entire bandwidth and downlink burst duration, exploiting both pilots and payloads pertaining to all \glspl{UE} using monostatic radar techniques. This allows achieving fairly good delay/range and Doppler resolution (less than 1 m and 5 km/h) at the cost of increased processing complexity, due to the addition of \gls{TF} sensing channel estimation, transformation into the \gls{DD} domain and delay-Doppler parameter extraction. It is worth noting that the monostatic \gls{BS} can only sense the absolute Doppler shift of targets, pertaining to the \gls{LoS} (direct) path. However, this absolute Doppler shift does not capture the residual Doppler spread experienced at the \gls{UE} side, which depends on the full geometric multipath configuration.
\begin{comment}
Although this information is useful for environment mapping, it is not sufficient to describe the residual Doppler spectrum experienced at the \gls{UE} side, which is the quantity of interest here, and it is a function of the geometric multipath configuration and \gls{UE} velocity, obtained after frequency synchronization at the \gls{UE} side, that aligns to the frequency of maximum energy (removing the absolute Doppler shift). 
\end{comment}
On the other hand, the sensing opportunities brought by each \gls{UE} are limited to Doppler spectrum (or spread) estimation. Indeed, each \gls{UE} is allocated with a finite bandwidth for a limited amount of time (in general, the resource scheduling is such that a \gls{UE} is allocated with resources for less than the downlink burst) thus only a coarse Doppler spectrum estimation is possible, to be fed back to the \gls{BS}. Moreover, since the \gls{UE} is not synchronized with the \gls{BS} and the bandwidth is scarce, it is not typically possible to locate the main scattering centers in space, which are \gls{UE}-specific and not observable by the monostatic \gls{BS}.

\subsubsection{Gear 2}
\gls{DD-a-OFDM} adds a flexible \gls{TF} pilot allocation to handle middle-to-high mobility conditions, expanding the range of application of Gear 1. From a mere sensing perspective, this does not alter the considerations made for Gear 1 with regard to the use of the \gls{BS} as a monostatic radar, since the emitted signal has similar \gls{DD} ambiguity properties and the processing chain is the same. Similarly, the addition of \gls{DD} processing at the \gls{UE} mainly brings advantages and disadvantages at the communication end (channel estimation and data decoding in mobility) but nothing changes for the Doppler spectrum estimation procedure, since each \gls{UE} operates on a different bandwidth or time slot, severely limiting the \gls{DD} resolution. 

\subsubsection{Gear 3}
\gls{DDW-OFDM} handles extreme mobility conditions by superposing a \gls{DD}-designed pilot signal to the \gls{TF} \gls{OFDM} waveform carrying data symbols. As the \gls{DD} pilot occupies much more bandwidth than the portion allocated to the single \gls{UE}, up to the entire available bandwidth and for the whole duration of the downlink burst, new opportunities for both \gls{BS} and \gls{UE} arise. A monostatic \gls{BS} exploring the surroundings using a properly designed \gls{DD} pilot will experience a much better range-Doppler ambiguity function, characterized by low sidelobes, without incurring additional computational complexity, since the sensing processing chain is the same. The improvement of range-Doppler ambiguity turns out to be beneficial for target detection in rich scattering environments with many close objects. Moreover, if the \gls{DD} pilot can expand outside of the nominal bandwidth limits---with a proper power downscaling to not generate harmful inter-cell interference---the range resolution can further increase, without compromising the \gls{UE} operation~\cite{Tagliaferri2024}. At the \gls{UE} side, instead, such \gls{DD} pilots allow having a refined Doppler spectrum estimation (thanks to the uniform pilot sampling in time resulting in an optimal Doppler ambiguity function) and a good multipath discrimination along range (thanks to the wide bandwidth).
%This dual-domain waveform concept unifies communication and sensing in a single transmitted signal.

% Communication performance
%TC:ignore
\begin{figure}[t]
 \centering
 \resizebox{0.99\columnwidth}{!}{
\begin{tikzpicture}

\begin{axis}[%
width=3.5in,
height=2in,
scale only axis,
xmin=0,
xmax=1000,
xtick={0, 100, 200, 300, 400, 500, 600, 700, 800, 900, 1000},
xticklabel style={
    /pgf/number format/set thousands separator={},
    /pgf/number format/1000 sep={}, % Utile per compatibilità con versioni meno recenti
    %font=\Large % Mantieni la dimensione del font che avevi impostato
},
xlabel={Speed (km/h)},
ymin=1,
ymax=2,
ylabel={Throughput [bits/RE]},
axis background/.style={fill=white},
axis x line*=bottom,
axis y line*=left,
xmajorgrids,
ymajorgrids,
legend style={at={(0.02,0.02)}, anchor=south west, legend cell align=left, align=left, draw=white!15!black, font=\scriptsize}
]

% --- DISEGNO DELLE BANDE COLORATE (BACKGROUND) ---
\fill[green!10] (axis cs:0,1) rectangle (axis cs:200,2);
\fill[yellow!10] (axis cs:200,1) rectangle (axis cs:600,2);
\fill[red!10] (axis cs:600,1) rectangle (axis cs:1000,2);

% --- ENTRATE LEGENDA PER LE BANDE ---
\addlegendimage{only marks, fill=green!10, draw=none, area legend}
\addlegendentry{Gear 1 Range}
\addlegendimage{only marks, fill=yellow!10, draw=none, area legend}
\addlegendentry{Gear 2 Range}
\addlegendimage{only marks, fill=red!10, draw=none, area legend}
\addlegendentry{Gear 3 Range}

% --- GEAR 1 ---
\addplot [color=red, line width=2pt]%, mark=square*, mark options={fill=red}]
table[row sep=crcr]{%
0 1.8202\\ 50 1.8199\\ 100 1.8179\\ 150 1.8117\\ 200 1.7980\\ 250 1.7705\\ 300 1.7025\\ 350 1.5917\\ 400 1.4797\\ 450 1.3649\\ 500 1.3054\\ 550 1.2671\\ 600 1.1878\\ 650 1.1663\\ 700 1.1508\\ 750 1.1336\\ 800 1.1230\\ 850 1.1124\\ 900 1.0889\\ 950 1.0948\\ 1000 1.0971\\
};
\addlegendentry{Classical OFDM}

% --- GEAR 2 (dt=4) ---
\addplot [color=blue, line width=2pt]%, mark=*, mark options={fill=blue}]
table[row sep=crcr]{%
0 1.8742\\ 50 1.8739\\ 100 1.8735\\ 150 1.8731\\ 200 1.8724\\ 250 1.8714\\ 300 1.8695\\ 350 1.7592\\ 400 1.6373\\ 450 1.5526\\ 500 1.4604\\ 550 1.4182\\ 600 1.3516\\ 650 1.3438\\ 700 1.2825\\ 750 1.2509\\ 800 1.2539\\ 850 1.2209\\ 900 1.2111\\ 950 1.1960\\ 1000 1.1698\\
};
\addlegendentry{DD-a-OFDM, $d_t=4$}

% --- GEAR 2 (dt=2) ---
\addplot [color=blue, dashed, line width=2pt]%, mark=triangle*, mark options={fill=white}]
table[row sep=crcr]{%
0 1.7490\\ 50 1.7490\\ 100 1.7486\\ 150 1.7483\\ 200 1.7476\\ 250 1.7468\\ 300 1.7456\\ 350 1.7442\\ 400 1.7428\\ 450 1.7410\\ 500 1.7398\\ 550 1.7373\\ 600 1.7349\\ 650 1.7092\\ 700 1.6192\\ 750 1.5546\\ 800 1.5176\\ 850 1.4502\\ 900 1.4233\\ 950 1.3909\\ 1000 1.3503\\
};
\addlegendentry{DD-a-OFDM, $d_t=2$}

% --- GEAR 3 (PDR=20) ---
\addplot [color=black, dashed, line width=2pt]%, mark=*, mark options={fill=black}]
table[row sep=crcr]{%
0 1.8293\\ 50 1.7993\\ 100 1.7997\\ 150 1.8009\\ 200 1.7985\\ 250 1.8017\\ 300 1.7959\\ 350 1.8028\\ 400 1.8094\\ 450 1.7999\\ 500 1.8064\\ 550 1.8031\\ 600 1.7994\\ 650 1.8061\\ 700 1.8101\\ 750 1.8044\\ 800 1.8064\\ 850 1.8021\\ 900 1.8066\\ 950 1.8086\\ 1000 1.8025\\
};
\addlegendentry{DDW-OFDM,  PDR$=20$ dB}

% --- GEAR 3 (PDR=25) ---
\addplot [color=black, line width=2pt]%, mark=triangle*, mark options={fill=black}]
table[row sep=crcr]{%
0 1.9088\\ 50 1.9071\\ 100 1.9114\\ 150 1.9131\\ 200 1.9112\\ 250 1.9145\\ 300 1.9121\\ 350 1.9150\\ 400 1.9188\\ 450 1.9216\\ 500 1.9190\\ 550 1.9217\\ 600 1.9192\\ 650 1.9164\\ 700 1.9200\\ 750 1.9227\\ 800 1.9213\\ 850 1.9201\\ 900 1.9254\\ 950 1.9247\\ 1000 1.9255\\
};
\addlegendentry{DDW-OFDM,  PDR$=25$ dB}

\end{axis}
\end{tikzpicture}}
\caption{Throughput comparison across the three gears as a function of the velocity of the scatterers. Simulation parameters consider a \gls{C-V2X} system operating at $5.9$~GHz with a \gls{SCS} of $15$~kHz. The channel follows an exponential \gls{PDP} with a rich-scattering Rayleigh fading model. For Gear~2 (\gls{DD-a-OFDM}), $d_t$ denotes the pilot spacing along the time axis, where a lower $d_t$ (higher pilot density) provides increased robustness against Doppler spread at the cost of spectral efficiency. The shaded regions indicate the suggested operational range for each gear: Gear~1 for low mobility (green), Gear~2 for medium mobility (yellow), and Gear~3 for high-mobility/high-reliability scenarios (red).} \label{fig:Throughput}
\end{figure}
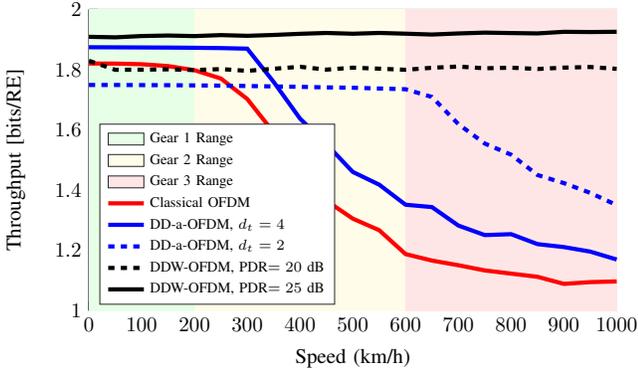

% PAPR - Gear 3
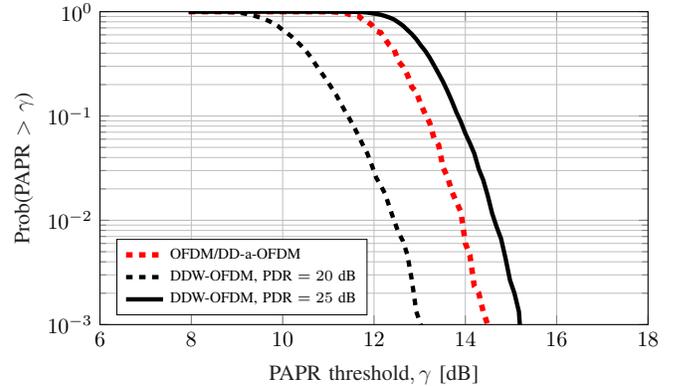
\begin{figure}[t]
 \centering
 \resizebox{\columnwidth}{!}{
\definecolor{mycolor1}{rgb}{1.00000,0.00000,1.00000}%
\definecolor{mycolor2}{rgb}{0.12941,0.12941,0.12941}%
\begin{tikzpicture}

\begin{axis}[%
width=3.5in,
height=2in,
at={(1.208in,0.616in)},
scale only axis,
xmin=6,
xmax=18,
xlabel style={font=\color{mycolor2}},
xlabel={$\text{PAPR threshold}, \gamma\text{ [dB]}$},
ymode=log,
ymin=0.001,
ymax=1,
yminorticks=true,
ylabel style={font=\color{mycolor2}},
ylabel={$\text{Prob(PAPR \textgreater{} }\gamma\text{)}$},
axis background/.style={fill=white},
title style={font=\bfseries\color{mycolor2}},
xmajorgrids,
ymajorgrids,
yminorgrids,
legend style={at={(0.03,0.03)}, anchor=south west, legend cell align=left, align=left, font=\scriptsize}
]
\addplot [color=red, dashed, line width=2.5pt]
  table[row sep=crcr]{%
7.94421519361183	1\\
8.04189603596327	1\\
8.13957687831471	1\\
8.23725772066615	1\\
8.3349385630176	1\\
8.43261940536904	1\\
8.53030024772048	1\\
8.62798109007192	1\\
8.72566193242336	1\\
8.82334277477481	1\\
8.92102361712625	1\\
9.01870445947769	1\\
9.11638530182913	1\\
9.21406614418058	1\\
9.31174698653202	1\\
9.40942782888346	1\\
9.5071086712349	1\\
9.60478951358635	1\\
9.70247035593779	1\\
9.80015119828923	1\\
9.89783204064067	1\\
9.99551288299211	1\\
10.0931937253436	1\\
10.190874567695	1\\
10.2885554100464	1\\
10.3862362523979	1\\
10.4839170947493	1\\
10.5815979371008	1\\
10.6792787794522	0.999666666666667\\
10.7769596218037	0.999666666666667\\
10.8746404641551	0.998666666666667\\
10.9723213065065	0.998333333333333\\
11.070002148858	0.998\\
11.1676829912094	0.990333333333333\\
11.2653638335609	0.983333333333333\\
11.3630446759123	0.979333333333333\\
11.4607255182637	0.942\\
11.5584063606152	0.926333333333333\\
11.6560872029666	0.915333333333333\\
11.7537680453181	0.835333333333333\\
11.8514488876695	0.802\\
11.949129730021	0.747\\
12.0468105723724	0.651333333333333\\
12.1444914147238	0.630666666666667\\
12.2421722570753	0.547333333333333\\
12.3398530994267	0.475666666666667\\
12.4375339417782	0.437333333333333\\
12.5352147841296	0.334\\
12.6328956264811	0.297\\
12.7305764688325	0.249\\
12.8282573111839	0.189666666666667\\
12.9259381535354	0.17\\
13.0236189958868	0.129666666666667\\
13.1212998382383	0.104333333333333\\
13.2189806805897	0.086\\
13.3166615229411	0.0606666666666667\\
13.4143423652926	0.0526666666666667\\
13.512023207644	0.032\\
13.6097040499955	0.026\\
13.7073848923469	0.0186666666666667\\
13.8050657346984	0.015\\
13.9027465770498	0.012\\
14.0004274194012	0.006\\
14.0981082617527	0.00466666666666667\\
14.1957891041041	0.00233333333333333\\
14.2934699464556	0.002\\
14.391150788807	0.00133333333333333\\
14.4888316311585	0.001\\
14.5865124735099	0.000333333333333333\\
14.6841933158613	0.000333333333333333\\
14.7818741582128	0.000333333333333333\\
14.8795550005642	0.000333333333333333\\
14.9772358429157	0\\
15.0749166852671	0\\
15.1725975276186	0\\
15.27027836997	0\\
15.3679592123214	0\\
15.4656400546729	0\\
15.5633208970243	0\\
15.6610017393758	0\\
15.7586825817272	0\\
15.8563634240786	0\\
15.9540442664301	0\\
16.0517251087815	0\\
16.149405951133	0\\
16.2470867934844	0\\
16.3447676358359	0\\
16.4424484781873	0\\
16.5401293205387	0\\
16.6378101628902	0\\
16.7354910052416	0\\
16.8331718475931	0\\
16.9308526899445	0\\
17.028533532296	0\\
17.1262143746474	0\\
17.2238952169988	0\\
17.3215760593503	0\\
17.4192569017017	0\\
17.5169377440532	0\\
17.6146185864046	0\\
};
\addlegendentry{OFDM/DD-a-OFDM}

\addplot [color=black,dashed, line width=2.0pt]
  table[row sep=crcr]{%
7.94421519361183	1\\
8.04189603596327	1\\
8.13957687831471	1\\
8.23725772066615	1\\
8.3349385630176	0.999666666666667\\
8.43261940536904	0.999666666666667\\
8.53030024772048	0.998666666666667\\
8.62798109007192	0.997\\
8.72566193242336	0.995666666666667\\
8.82334277477481	0.993666666666667\\
8.92102361712625	0.987666666666667\\
9.01870445947769	0.981\\
9.11638530182913	0.967666666666667\\
9.21406614418058	0.95\\
9.31174698653202	0.929\\
9.40942782888346	0.904333333333333\\
9.5071086712349	0.878666666666667\\
9.60478951358635	0.847666666666667\\
9.70247035593779	0.812666666666667\\
9.80015119828923	0.771333333333333\\
9.89783204064067	0.727333333333333\\
9.99551288299211	0.671\\
10.0931937253436	0.626666666666667\\
10.190874567695	0.579333333333333\\
10.2885554100464	0.528\\
10.3862362523979	0.480666666666667\\
10.4839170947493	0.433666666666667\\
10.5815979371008	0.384333333333333\\
10.6792787794522	0.332333333333333\\
10.7769596218037	0.287666666666667\\
10.8746404641551	0.250333333333333\\
10.9723213065065	0.218333333333333\\
11.070002148858	0.186666666666667\\
11.1676829912094	0.160666666666667\\
11.2653638335609	0.134\\
11.3630446759123	0.112\\
11.4607255182637	0.0946666666666667\\
11.5584063606152	0.079\\
11.6560872029666	0.0653333333333333\\
11.7537680453181	0.053\\
11.8514488876695	0.044\\
11.949129730021	0.034\\
12.0468105723724	0.0256666666666667\\
12.1444914147238	0.0216666666666667\\
12.2421722570753	0.018\\
12.3398530994267	0.0133333333333333\\
12.4375339417782	0.01\\
12.5352147841296	0.00766666666666667\\
12.6328956264811	0.00633333333333333\\
12.7305764688325	0.00466666666666667\\
12.8282573111839	0.00266666666666667\\
12.9259381535354	0.00133333333333333\\
13.0236189958868	0.001\\
13.1212998382383	0.000666666666666667\\
13.2189806805897	0\\
13.3166615229411	0\\
13.4143423652926	0\\
13.512023207644	0\\
13.6097040499955	0\\
13.7073848923469	0\\
13.8050657346984	0\\
13.9027465770498	0\\
14.0004274194012	0\\
14.0981082617527	0\\
14.1957891041041	0\\
14.2934699464556	0\\
14.391150788807	0\\
14.4888316311585	0\\
14.5865124735099	0\\
14.6841933158613	0\\
14.7818741582128	0\\
14.8795550005642	0\\
14.9772358429157	0\\
15.0749166852671	0\\
15.1725975276186	0\\
15.27027836997	0\\
15.3679592123214	0\\
15.4656400546729	0\\
15.5633208970243	0\\
15.6610017393758	0\\
15.7586825817272	0\\
15.8563634240786	0\\
15.9540442664301	0\\
16.0517251087815	0\\
16.149405951133	0\\
16.2470867934844	0\\
16.3447676358359	0\\
16.4424484781873	0\\
16.5401293205387	0\\
16.6378101628902	0\\
16.7354910052416	0\\
16.8331718475931	0\\
16.9308526899445	0\\
17.028533532296	0\\
17.1262143746474	0\\ 
17.2238952169988	0\\
17.3215760593503	0\\
17.4192569017017	0\\
17.5169377440532	0\\
17.6146185864046	0\\
};
\addlegendentry{DDW-OFDM, PDR $=20$ dB}

\addplot [color=black, line width=2.0pt]
  table[row sep=crcr]{%
7.94421519361183	1\\
8.04189603596327	1\\
8.13957687831471	1\\
8.23725772066615	1\\
8.3349385630176	1\\
8.43261940536904	1\\
8.53030024772048	1\\
8.62798109007192	1\\
8.72566193242336	1\\
8.82334277477481	1\\
8.92102361712625	1\\
9.01870445947769	1\\
9.11638530182913	1\\
9.21406614418058	1\\
9.31174698653202	1\\
9.40942782888346	1\\
9.5071086712349	1\\
9.60478951358635	1\\
9.70247035593779	1\\
9.80015119828923	1\\
9.89783204064067	1\\
9.99551288299211	1\\
10.0931937253436	1\\
10.190874567695	1\\
10.2885554100464	1\\
10.3862362523979	1\\
10.4839170947493	1\\
10.5815979371008	1\\
10.6792787794522	1\\
10.7769596218037	1\\
10.8746404641551	0.999666666666667\\
10.9723213065065	0.999666666666667\\
11.070002148858	0.999666666666667\\
11.1676829912094	0.999666666666667\\
11.2653638335609	0.999666666666667\\
11.3630446759123	0.999\\
11.4607255182637	0.998\\
11.5584063606152	0.997\\
11.6560872029666	0.994333333333333\\
11.7537680453181	0.990333333333333\\
11.8514488876695	0.982333333333333\\
11.949129730021	0.966666666666667\\
12.0468105723724	0.953333333333333\\
12.1444914147238	0.935333333333333\\
12.2421722570753	0.909\\
12.3398530994267	0.878333333333333\\
12.4375339417782	0.842333333333333\\
12.5352147841296	0.789\\
12.6328956264811	0.735\\
12.7305764688325	0.669\\
12.8282573111839	0.609666666666667\\
12.9259381535354	0.545\\
13.0236189958868	0.476\\
13.1212998382383	0.421333333333333\\
13.2189806805897	0.362333333333333\\
13.3166615229411	0.304666666666667\\
13.4143423652926	0.255\\
13.512023207644	0.212666666666667\\
13.6097040499955	0.172666666666667\\
13.7073848923469	0.138\\
13.8050657346984	0.108\\
13.9027465770498	0.0876666666666667\\
14.0004274194012	0.068\\
14.0981082617527	0.055\\
14.1957891041041	0.0436666666666667\\
14.2934699464556	0.031\\
14.391150788807	0.0243333333333333\\
14.4888316311585	0.0173333333333333\\
14.5865124735099	0.0116666666666667\\
14.6841933158613	0.00866666666666667\\
14.7818741582128	0.00666666666666667\\
14.8795550005642	0.00433333333333333\\
14.9772358429157	0.00266666666666667\\
15.0749166852671	0.002\\
15.1725975276186	0.00133333333333333\\
15.27027836997	0.000333333333333333\\
15.3679592123214	0\\
15.4656400546729	0\\
15.5633208970243	0\\
15.6610017393758	0\\
15.7586825817272	0\\
15.8563634240786	0\\
15.9540442664301	0\\
16.0517251087815	0\\
16.149405951133	0\\
16.2470867934844	0\\
16.3447676358359	0\\
16.4424484781873	0\\
16.5401293205387	0\\
16.6378101628902	0\\
16.7354910052416	0\\
16.8331718475931	0\\
16.9308526899445	0\\
17.028533532296	0\\
17.1262143746474	0\\
17.2238952169988	0\\
17.3215760593503	0\\
17.4192569017017	0\\
17.5169377440532	0\\
17.6146185864046	0\\
};
\addlegendentry{DDW-OFDM, PDR $=25$ dB}

\end{axis}

\end{tikzpicture}%
}
 \caption{\gls{PAPR} characterization of the gears. The \gls{CCDF} curves show the probability of the instantaneous \gls{PAPR} exceeding a given threshold $\gamma$ for Gear~1/2 (\gls{OFDM}/\gls{DD-a-OFDM}) and Gear~3 at \gls{PDR} levels of $20$~dB and $25$~dB.} \label{fig:PAPRfigures}
\end{figure}
%TC:endignore

\subsection{Performance Evaluation}
To highlight the necessity of the multi-gear framework, Fig. \ref{fig:Throughput} shows the communication throughput, expressed in terms of bits per \gls{RE}, for all the waveform gears. 
Gear 1 (legacy \gls{OFDM}) achieves high throughput in low-mobility regimes (up to approximately $200$ km/h) due to its minimal pilot overhead and symbol-level processing. However, as the Doppler spread increases, its performance degrades sharply, becoming unreliable for high-speed scenarios.
In this context, Gear 2 (\gls{DD-a-OFDM}) serves as a crucial bridge. By employing \gls{DD} domain channel estimation while maintaining the standard \gls{TF} pilot structure, it significantly extends the reliable operating range without requiring changes to the \gls{3GPP} transmitter-side standardization. As seen in Fig. \ref{fig:Throughput}, Gear 2 maintains a robust throughput in medium mobility range where Gear 1 fails. The processing burden is left to the receiver, allowing operators to support reliable communications at high speeds using the existing infrastructure.
For high mobility (above $600$ km/h), Gear 3 (\gls{DDW-OFDM}) becomes indispensable. While it requires a new pilot design and higher receiver complexity, it provides a near-constant throughput across the 6G mobility range, regardless of the velocity. The choice of the \gls{PDR} is a key trade-off: a \gls{PDR} of $25$ dB offers the highest reliability regardless of the mobility conditions, whereas a \gls{PDR} of $20$ dB provides performance comparable to legacy \gls{OFDM} at lower speeds but with a significantly reduced impact on the \gls{PAPR}, as shown by Fig. \ref{fig:PAPRfigures}. The selection of the optimal \gls{PDR} represents a fundamental design step, as it serves as a key tuning knob to maximize throughput while precisely meeting the target \gls{PAPR} requirements \cite{marchese2025robust6gofdmhighmobility}.

\begin{comment}
These results justify the \gls{GS-OFDM} strategy as no single processing strategy is optimal across the entire mobility range:
Gear 1 is used for the vast majority of users to ensure low latency and maximum power efficiency. Gear 2 is activated as a bridge to maintain connectivity at high speeds, leveraging backward compatibility. Gear 3 is reserved only for extreme mobility, where the reliability requirement justifies the increased receiver complexity and new standardization demands. 
\end{comment}
A summary of the characteristics of the different waveform gears is provided in Table \ref{tab:summary}.

%TC:ignore
\begin{table*}[!t]
\centering
\caption{Summary of the three processing gears in the \gls{GS-OFDM} framework.}
\label{tab:summary}
\footnotesize
\begin{tabular}{@{}l p{3.8cm} p{3.8cm} p{3.8cm}@{}}
\toprule
\textbf{Property} & \textbf{Gear~1: legacy OFDM} & \textbf{Gear~2: DD-a-OFDM} & \textbf{Gear~3: DDW-OFDM} \\
\midrule
Mobility regime & Quasi-static & Time-varying & Fast-varying \\
\addlinespace
TX pilot scheme & TF embedded & TF embedded (possibly denser) & DD superimposed pilot \\
\addlinespace
Channel estimation & TF interpolation & DD parameter estimation via  periodic CSF & DD parameter estimation \\
\addlinespace
Equalization & Single-tap (TF domain) & MMSE or IMFC & IMFC \\
\addlinespace
Processing latency & OFDM symbol & OFDM frame & OFDM frame \\
\addlinespace
UE processing burden & Low & Medium & High \\
\addlinespace
Pilot overhead & Standard NR overhead & Standard or slightly increased & No RE loss due to power sharing \\
\addlinespace
PAPR & Standard & Standard & Tunable via PDR selection \\
\addlinespace
Monostatic sensing at BS & Basic range-Doppler map & Same as Gear 1 & Enhanced range-Doppler ambiguity, higher DD resolution \\
\addlinespace
5G NR compatibility & Full & Largely compatible & Requires new pilot design and standardization \\
\bottomrule
\end{tabular}
\end{table*}
%TC:endignore

%% ====================================================================
%\section{Outlook and Research Challenges}
\section{Opportunities and Open Challenges}
\label{sec:outlook}

This section highlights interesting opportunities arising from the use of \gls{GS-OFDM} and related open challenges, calling for dedicated research efforts.
%% ====================================================================

\subsection{Impact of MIMO and Spatial Filtering on Gear Switching}\label{sec:impactMIMO}
The adoption of \gls{MIMO} and spatial filtering acts as a fundamental performance multiplier across all gears, primarily by reshaping the perceived channel dynamics.
The Doppler spread experienced by the receiver is intrinsically linked to the angular spread of the multipath components. In systems with large antenna arrays (particularly at FR2 and sub-THz), narrow-beam spatial filtering attenuates paths outside the main lobe. This reduces the effective angular spread and, consequently, the effective Doppler spread.
A direct consequence of this spatial cleaning is that the coherence time is effectively extended. Therefore, Gear~1 can remain reliable up to velocity values where it would otherwise fail under isotropic multipath conditions. This effect shifts the switching points between gears: with high-gain beamforming, the transition to Gear~2 or Gear~3 occurs at higher mobility levels compared to the levels presented in Table~\ref{tab:coherence}.

\subsection{Handling Heterogeneous Intra-Cluster Mobility}
Upon collecting Doppler spectrum or spread estimates from the \glspl{UE}, the \gls{BS} performs user clustering based on spatial proximity and mobility profiles. A significant challenge arises when co-located users experience heterogeneous channel dynamics: the \gls{BS} may adopt a worst-case strategy (selecting the gear required by the user with the highest experienced mobility) or try to balance rate-energy trade-off (choosing the gear that best balances throughput and \gls{UE} energy consumption across the cluster). The implementation of per-\gls{UE} dedicated gear selection remains an open research frontier, requiring advances in multi-user precoding to enable fully individualized waveform adaptation.

\subsection{Operating UEs over Wide Bandwidths for Gear 3}
As discussed above, Gear 3 calls for the use of a \gls{DD}-designed pilot superimposed to the entire \gls{TF} resource grid used for the compound \gls{OFDM} downlink signal burst. This implies that \glspl{UE} in extreme mobility conditions shall operate the receiving circuitry (analog and digital) on the full available bandwidth (in the order of hundreds of MHz) instead of operating only on the allocated bandwidth where the intended data payload is transmitted, in order to receive and process the \gls{DD} pilot. Therefore, the energy expenditure of the \gls{UE}, mainly due to digital sampling and processing, increases considerably. On one side, this may be regarded as the cost of extreme mobility in order to support high-data rate communications (see Fig. \ref{fig:Throughput}), but it is also open for the investigation on the sufficient duty cycle for the transmission of the \gls{DD} pilot by the \gls{BS}, as a trade-off between communication (and sensing) performance and energy consumption. 

\subsection{Cooperative ISAC-assisted Doppler Spread Prediction}
A promising extension of \gls{GS-OFDM} is to exploit cooperative \gls{ISAC} to construct and continuously update a geometry-aware radio environment map. By fusing sensing observations from multiple \glspl{BS} or network nodes, the infrastructure can localize dominant scatterers, infer their motion, and predict the Doppler support likely to be experienced by each \gls{UE} along its trajectory. Such network-side situational awareness can provide priors for gear selection, switching-threshold adaptation, and Doppler-aware scheduling, thereby reducing the overhead by eliminating the need for continuous Doppler-spread feedback from the \gls{UE} to the \gls{BS}. In this vision, the \gls{BS} can autonomously estimate the channel dynamics by collecting information from the broader sensing fabric. Nevertheless, complete feedback elimination should not be assumed in general, since the residual Doppler spread after local synchronization remains user- and beam-dependent and may still require occasional \gls{UE}-side reporting or calibration.

\subsection{Uplink Gear Design}
While this work focuses on the downlink, the definition of uplink processing gears remains an open research problem. A natural starting point could be to extend the gear concept to \gls{SC-FDMA}, the baseline uplink waveform confirmed for \gls{6G}, by equipping it with \gls{DD}-domain channel estimation capabilities analogous to those of Gears 2 and 3. Alternatively, \gls{DDW-OFDM} represents a compelling uplink candidate: by appropriately selecting the \gls{PDR}, it can achieve a \gls{PAPR} lower than that of legacy \gls{OFDM}, while providing near-constant throughput regardless of the Doppler spread. However, extending \gls{DDW-OFDM} to a multi-user uplink setting introduces non-trivial challenges. Since each \gls{UE} must superimpose a \gls{DD}-domain pilot over the entire bandwidth and frame duration, the available \gls{DD} resources must be partitioned among simultaneous uplink users, reducing the \gls{DD}-domain resolution achievable at the \gls{BS} for each individual link. Furthermore, requiring each \gls{UE} to transmit over the full bandwidth significantly increases device energy consumption. 

\section{Conclusion}
In this article, we have envisioned a novel \gls{GS-OFDM} framework that adaptively selects among three processing gears to address the critical challenge of maintaining reliable communications in high-mobility \gls{6G} scenarios, where traditional \gls{OFDM} systems suffer from severe \gls{ICI}. Our solution enables a seamless transition for integrating \gls{DD} processing into existing infrastructures, ensuring robust communication performance across the entire range of mobility supported by \gls{6G}.
Furthermore, we have discussed the integration of this multi-gear approach with \gls{MIMO} beamforming and \gls{ISAC} capabilities, highlighting how spatial filtering and sensing can further enhance spectral efficiency and reduce system latency. Finally, we have outlined the key research opportunities and implementation challenges that must be addressed to pave the way for \gls{GS-OFDM} operation in future 6G networks.

\ifCLASSOPTIONcaptionsoff
  \newpage
\fi

%\balance
%TC:ignore
\bibliographystyle{IEEEtran}
\bibliography{reference}
%TC:endignore

\end{document}